\documentclass[conference]{IEEEtran}
\IEEEoverridecommandlockouts

\usepackage[backend=bibtex]{biblatex}
\usepackage{amsmath,amssymb,amsfonts}
\usepackage{graphicx}
\usepackage{textcomp}
\usepackage{color}
\usepackage[mathcal]{eucal}
\usepackage[utf8]{inputenc}
\usepackage[margin=1in]{geometry}
\usepackage{hhline}
\usepackage{anyfontsize}
\usepackage{bigdelim}
\usepackage{xspace}
\usepackage{etoolbox}
\AtBeginEnvironment{algorithm}{\linespread{1.2}\selectfont}
\AtBeginEnvironment{tikzpicture}{\linespread{1.2}\selectfont}
\AtBeginEnvironment{figure}{\linespread{1.2}\selectfont}
\AtBeginEnvironment{enumerate}{\linespread{1.2}\selectfont}
\AtBeginEnvironment{description}{\linespread{1.2}\selectfont}
\AtBeginEnvironment{itemize}{\linespread{1.2}\selectfont}
\AtBeginEnvironment{caption}{\linespread{1}\selectfont}
\usepackage{subfigure}
\usepackage{framed}
\usepackage{collcell}
\usepackage{cancel} 
\usepackage{multicol}
\usepackage{multirow}
\usepackage[dvipsnames]{xcolor}
\usepackage{colortbl}
\usepackage{tikz}
\usetikzlibrary{tikzmark,decorations.pathreplacing,matrix,calc,arrows,positioning,backgrounds,shapes,trees,shadows}
\usepackage{pgfplots}
\pgfplotsset{compat=1.13}
\usepackage[linesnumbered,vlined,ruled]{algorithm2e}
\usepackage{listings}
\usepackage{fancyvrb}
\usepackage{url}
\usepackage{hyperref}
\usepackage[capitalise]{cleveref}

\def\BibTeX{{\rm B\kern-.05em{\sc i\kern-.025em b}\kern-.08em T\kern-.1667em\lower.7ex\hbox{E}\kern-.125emX}}

\newcommand{\ttt}[1]{\texttt{#1}}
\newcommand{\reals}{\mathbb{R}}
\newcommand{\realsn}{\mathbb{R}^n}

\newcommand{\realsm}{\mathbb{R}^m}

\newcommand{\transpose}{^{\scriptscriptstyle \top}}
\newcommand{\flo}[1]{\left\lfloor{#1}\right\rfloor}
\newcommand{\cel}[1]{\left\lceil{#1}\right\rceil}

\newcommand{\trm}[1]{\textrm{#1}}
\newcommand{\verteq}{\rotatebox{90}{$\,=$}}

\DeclareMathOperator{\bigo}{O}

\begin{document}

\title{Efficient Distributed-Memory Parallel Matrix-Vector Multiplication with
       Wide or Tall Unstructured Sparse Matrices\\
\thanks{This research was supported in part by National Science 
        Foundation grant CCF-1115638.}
}

\author{\IEEEauthorblockN{Jonathan Eckstein}
\IEEEauthorblockA{\textit{MSIS Department and RUTCOR}\\
\textit{Rutgers University}\\
Piscataway, NJ, 08854\\
jeckstei@business.rutgers.edu}
\and
\IEEEauthorblockN{Gy\"orgy M\'aty\'asfalvi}
\IEEEauthorblockA{\textit{CS and Math}\\
\textit{Brookhaven National Lab}\\
Upton, NY, 11973\\
gmatyasfalvi@bnl.gov}
}

\maketitle

\pdfoutput=1

\begin{abstract}
This paper presents an efficient technique for matrix-vector and
vector-transpose-matrix multiplication in distributed-memory parallel
computing environments, where the matrices are unstructured, sparse, and have
a substantially larger number of columns than rows or \emph{vice versa}. Our
method allows for parallel I/O, does not require extensive preprocessing, and
has the same communication complexity as matrix-vector multiplies with column
or row partitioning. Our implementation of the method uses MPI.
We partition the matrix by individual nonzero elements, rather than by row or
column, and use an ``overlapped'' vector representation that is matched to the
matrix.  The transpose multiplies use matrix-specific MPI communicators
and reductions that we show can be set up in an efficient manner. The proposed
technique achieves a good work per processor balance even if some of the
columns are dense, while keeping communication costs relatively low.
\end{abstract}

\begin{IEEEkeywords}
Linear algebra, Sparse matrices, Parallel algorithms, Distributed algorithms
\end{IEEEkeywords}

\section{Introduction and Motivation}
\label{SEC:Intro}

The sparse matrix-vector multiply (SpMV) operation multiplies a sparse
matrix $A \in \reals^{m \times n}$  by a dense vector $x \in \realsn$,
resulting in a dense vector $y \in \realsm$, $Ax = y$. The sparse
vector-transpose-matrix (SpVTM) operation multiplies $A$ by the
transpose of a dense vector $v \in \realsm$ resulting in the transpose of a
dense vector $u \in \realsn$, $v\transpose A =  u\transpose$. The 
sequential implementation of these operations is straightforward and possibly
easily parallelizable on shared-memory computers, for example, using
\ttt{pragma} directives~\cite{MG:Pragma}. However, implementing large-scale SpMVs
and SpVTMs on distributed-memory supercomputers requires careful
attention to detail. Various approaches exist if $A$ has some structure, for
example, if it is block-diagonal~\cite{MG:Vuduc05, MG:Tuminaro95}. If $A$ lacks such structure, standard row or column
partitioning techniques may have difficulty producing an even load balance, or
may require extensive preprocessing of $A$ in order to do so~\cite{MG:Catalyurek99, MG:Buluc09}. For example,
some document-term matrices~\cite{MG:Dtmtx} arising in text classification
problems~\cite{MG:TextClass} are sparse and wide, \emph{i.e.} have $m \ll n$, and contain
a small number of highly dense columns. A natural way of partitioning wide
matrices is to assign subsets of the columns to individual processors.
However, this approach may result in a very uneven load balance in the
presence of dense columns. This paper describes a nonzero partitioning
technique that avoids this issue by assigning subsets of nonzero matrix
entries, instead of columns, to individual processors, and adjusting the
representation of vectors to match the representation of the matrix.

\section{Data Distribution}
\label{SEC:Dist}
For the sake of brevity, we will assume throughout this paper that $A$ has
many more columns than rows, that is, $m \ll n$, and is stored in column major
order~\cite{MG:SparseStorage}. By reversing the roles of rows and columns, our
method may be applied analogously to ``tall'' matrices with $m \gg n$,  stored
in row major order. Our partitioning method distributes the nonzero
elements of $A$ evenly between processors and can result in
one column of $A$ being ``owned'' by multiple processors; the details will be
covered below. Coefficients of vectors of length $n$, such as $x$ and $u$
above, are also distributed across processors.  On the other hand, we
replicate vectors of length $m$, such as $y$ and $v$ above, in the memory of
all processors. Since these vectors are much shorter than $x$ or $u$, the
memory usage impact of this replication is limited. Throughout the
discussion, we treat each processor as having its own memory; this does not
prevent the method from being implemented on a system in which physical memory
is (partially) shared.

Suppose that there are $P$ processors denoted $p_0$ through $p_{P-1}$, and let
$\{J_0,\ldots,J_{P-1}\}$ be a collection of subsets of $\{1,\ldots,n\}$, with
$J_i$ denoting the set of column indices assigned to 
$p_i$. An index
$j \in \{1,\ldots,n\}$ might be assigned to multiple $J_i$, hence,
$\{J_0,\ldots,J_{P-1}\}$ is a cover and in general not a partition
of $\{1,\ldots,n\}$. 
We momentarily defer discussing how we determine the cover
$\{J_0,\ldots,J_{P-1}\}$. Let $x_{(i)}$ and $u_{(i)}$ denote the respective
subvectors of $x$ and $u$ consisting of the coefficients with indices in
$J_i$.
~

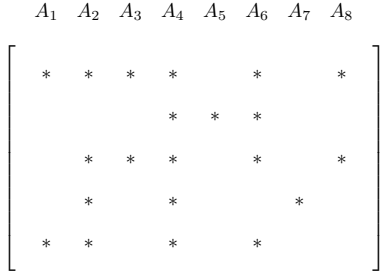
\begin{figure}
\begin{center}
\scalebox{0.75}{
\begin{tikzpicture}[
bmatrix/.style={matrix of math nodes, nodes in empty cells, text width=1.5em, text height=1em, text depth=.5em, row sep=-\pgflinewidth, column sep=-\pgflinewidth, ampersand replacement=\&, right delimiter=\rbrack, left delimiter= \lbrack}, 
pmatrix/.style={matrix of math nodes, nodes in empty cells, text width=1.5em, text height=1em, text depth=.5em, row sep=-\pgflinewidth, column sep=-\pgflinewidth, ampersand replacement=\&}
]

\matrix(ColLabel)[pmatrix, nodes={anchor=center, rectangle, align=center}] {
A_1 \&  A_2  \& A_3  \& A_4 \& A_5 \& A_6 \& A_7 \& A_8 \\ };

\matrix(A)[bmatrix, below=.25em of ColLabel, nodes={anchor=center, rectangle, align=center}] {
* \&  *  \& *  \& * \&   \& * \&   \& * \\ 
  \&     \&    \& * \& * \& * \&   \&   \\
  \&  *  \& *  \& * \&   \& * \&   \& * \\
  \&  *  \&    \& * \&   \&   \& * \&   \\
* \&  *  \&    \& * \&   \& * \&   \&   \\};
\end{tikzpicture}}
\end{center}
\vspace{-2.5ex}
\caption{A sparse Matrix, its nonzeros indicated by *.}\label{FIG:SparseMtx}
\end{figure}

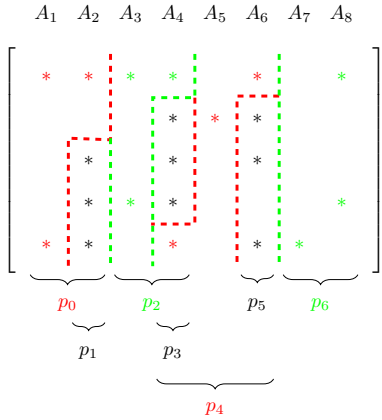
\begin{figure}
\begin{center}
\scalebox{0.75}{
\begin{tikzpicture}[
bmatrix/.style={matrix of math nodes, nodes in empty cells, text width=1.5em, text height=1em, text depth=.5em, row sep=-\pgflinewidth, column sep=-\pgflinewidth, ampersand replacement=\&, right delimiter=\rbrack, left delimiter= \lbrack}, 
pmatrix/.style={matrix of math nodes, nodes in empty cells, text width=1.5em, text height=1em, text depth=.5em, row sep=-\pgflinewidth, column sep=-\pgflinewidth, ampersand replacement=\&}
]

\matrix(ColLabel)[pmatrix, nodes={anchor=center, rectangle, align=center}] {
A_1 \&  A_2  \& A_3  \& A_4 \& A_5 \& A_6 \& A_7 \& A_8 \\ };

\matrix(A)[bmatrix, below=.25em of ColLabel, nodes={anchor=center, rectangle, align=center}] {
|[red]|*  \&|[red]|  * 	\&|[green]| *	\&|[green]| *	\&	    \&|[red]| * \&	   	\& |[green]| *  \\ 
	  \&		\&		\& *		\&|[red]| * \& * 	\&   		\&   		\\
	  \& * 		\&		\& * 		\& 	    \& * 	\&   		\&   		\\
	  \& * 		\&|[green]| *   \& * 		\& 	    \&   	\&   		\& |[green]| *  \\
|[red]|*  \& * 		\&    		\&|[red]| * 	\& 	    \& * 	\&|[green]|* 	\&   		\\};
	  
\draw[dashed, red, line width=1.5pt] (A-1-2.north east) -- (A-2-2.south east) -- (A-3-1.north east) -- (A-5-1.south east);
\draw[dashed, green, line width=1.5pt] (A-3-2.north east) -- (A-5-2.south east);
\draw[dashed, green, line width=1.5pt] (A-1-4.north east) -- (A-1-4.south east) -- (A-1-3.south east) -- (A-5-3.south east);
\draw[dashed, red, line width=1.5pt] (A-1-4.south east) -- (A-4-4.south east) -- (A-4-4.south west);
\draw[dashed, red, line width=1.5pt] (A-2-6.north east) -- (A-2-5.north east) -- (A-5-5.south east);
\draw[dashed, green, line width=1.5pt] (A-1-6.north east) -- (A-5-6.south east);
\draw[decorate,decoration={brace,mirror,amplitude=5pt}]($(A-5-1.south west)+(.1,-.15)$) -- ($(A-5-2.south east)+(-.1,-.15)$) node [red,midway,yshift=-15pt]{$p_0$};
\draw[decorate,decoration={brace,mirror,amplitude=4pt}]($(A-5-2.south west)+(.1,-1)$) -- ($(A-5-2.south east)+(-.1,-1)$) node [midway,yshift=-15pt]{$p_1$};
\draw[decorate,decoration={brace,mirror,amplitude=5pt}]($(A-5-3.south west)+(.1,-.15)$) -- ($(A-5-4.south east)+(-.1,-.15)$) node [green,midway,yshift=-15pt]{$p_2$};
\draw[decorate,decoration={brace,mirror,amplitude=4pt}]($(A-5-4.south west)+(.1,-1)$) -- ($(A-5-4.south east)+(-.1,-1)$) node [midway,yshift=-15pt]{$p_3$};
\draw[decorate,decoration={brace,mirror,amplitude=5pt}]($(A-5-4.south west)+(.1,-2)$) -- ($(A-5-6.south east)+(-.1,-2)$) node [red,midway,yshift=-15pt]{$p_4$};
\draw[decorate,decoration={brace,mirror,amplitude=4pt}]($(A-5-6.south west)+(.1,-.15)$) -- ($(A-5-6.south east)+(-.1,-.15)$) node [midway,yshift=-15pt]{$p_5$};
\draw[decorate,decoration={brace,mirror,amplitude=5pt}]($(A-5-7.south west)+(.1,-.15)$) -- ($(A-5-8.south east)+(-.1,-.15)$) node [green,midway,yshift=-15pt]{$p_6$};
\end{tikzpicture}}
\end{center}
\vspace{-3.5ex}
\caption{Even partitioning of nonzeros between
         processors.}\label{FIG:NonzeroPart}
\end{figure}

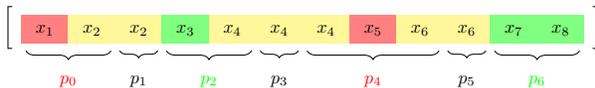
\begin{figure}
\begin{center}
\scalebox{0.675}{
\begin{tikzpicture}[
bvector/.style={matrix of math nodes, nodes in empty cells, text depth=.5ex, text height=1.5ex, text width=2em, row sep=-\pgflinewidth, column sep=-\pgflinewidth, ampersand replacement=\&, right delimiter=\rbrack, left delimiter= \lbrack} ]

\matrix(x)[bvector, nodes={anchor=base, rectangle, align=center}] {
|[fill=red!50]| x_1  \& |[fill=yellow!50]| x_2  \& |[fill=yellow!50]| x_2 \& |[fill=green!50]| x_3 \& |[fill=yellow!50]| x_4  \& |[fill=yellow!50]| x_4  \& |[fill=yellow!50]| x_4 \& |[fill=red!50]| x_5  \& |[fill=yellow!50]| x_6  \& |[fill=yellow!50]| x_6  \& |[fill=green!50]| x_7 \& |[fill=green!50]| x_8 \\ };

\draw[decorate,decoration={brace,mirror,amplitude=5pt}]($(x-1-1.south west)+(.1,-.15)$) -- ($(x-1-2.south east)+(-.1,-.15)$) node [red,midway,yshift=-17pt]{$p_0$};
\draw[decorate,decoration={brace,mirror,amplitude=4pt}]($(x-1-3.south west)+(.1,-.15)$) -- ($(x-1-3.south east)+(-.1,-.15)$) node [midway,yshift=-17pt]{$p_1$};
\draw[decorate,decoration={brace,mirror,amplitude=5pt}]($(x-1-4.south west)+(.1,-.15)$) -- ($(x-1-5.south east)+(-.1,-.15)$) node [green,midway,yshift=-17pt]{$p_2$};
\draw[decorate,decoration={brace,mirror,amplitude=4pt}]($(x-1-6.south west)+(.1,-.15)$) -- ($(x-1-6.south east)+(-.1,-.15)$) node [midway,yshift=-17pt]{$p_3$};
\draw[decorate,decoration={brace,mirror,amplitude=5pt}]($(x-1-7.south west)+(.1,-.15)$) -- ($(x-1-9.south east)+(-.1,-.15)$) node [red,midway,yshift=-17pt]{$p_4$};
\draw[decorate,decoration={brace,mirror,amplitude=4pt}]($(x-1-10.south west)+(.1,-.15)$) -- ($(x-1-10.south east)+(-.1,-.15)$) node [midway,yshift=-17pt]{$p_5$};
\draw[decorate,decoration={brace,mirror,amplitude=5pt}]($(x-1-11.south west)+(.1,-.15)$) -- ($(x-1-12.south east)+(-.1,-.15)$) node [green,midway,yshift=-17pt]{$p_6$};
\end{tikzpicture}}
\end{center}
\vspace{-2.5ex}
\caption{Representation of $x$ given the nonzero partition in~\cref{FIG:NonzeroPart}.}\label{FIG:xCoefReplication}
\end{figure}

Let $Z$ denote the total number of nonzero entries in $A$, and for each
$i=0,\ldots,P-1$, let $A_{(i)}$ denote the ``local matrix'' consisting of the
nonzero elements assigned to processor $p_i$. We determine  the submatrices
$\{A_{(i)}\}$ by dividing the $Z$ nonzeros of $A$ into $P$ groups, each
contiguous within the row-major order, whose size varies by at most one
element: formally, we may take the first $Z\, (\!\!\!\!\mod P)$ 
groups to have size
$\cel{\frac{Z}{P}}$ and the remaining groups to have size $\flo{\frac{Z}{P}}$.
This partitioning then determines each cover element $J_i$, which consists of
those columns $j$ for which $A_{(i)}$ contains a nonzero in column $j$.
\cref{FIG:SparseMtx,FIG:NonzeroPart} depict an example of such a partitioning,
spreading the $21$ nonzeros of an $8$-column matrix across $7$ processors,
leaving each processor with exactly $3$ nonzeros (this matrix 
does not have $m \ll n$, but we still use it for purposes of illustration). As
\cref{FIG:NonzeroPart} suggests, this technique can assign parts of the same
column to different processors; if a local matrix $A_{(i)}$ has a partial
column, then the missing elements are locally treated as being zero. For
example, consider the first two local matrices in~\cref{FIG:NonzeroPart},
$A_{(0)}$ and $A_{(1)}$. The first of these local matrices, $A_{(0)}$, is
stored in $p_0$ and is represented as a sparse matrix with $2$
columns, having $2$ nonzeros in the first column and $1$ nonzero in the
second.  The second local matrix, $A_{(1)}$, is stored in 
$p_1$ and is represented as a sparse matrix with $1$ column containing $3$
nonzeros. Thus, global column $A_2$ is split between processors $p_0$ and
$p_1$.  As a consequence, both $J_0$ and $J_1$ contain the index $2$;
specifically, $J_0 = \{1,2\}$ and $J_1 = \{2\}$.


Whenever two or more processors are responsible for one column, we call that
column's index an \emph{overlap zone}.  If index $j$ is an overlap zone, then
global column $A_j$ is the sum of the corresponding columns stored on each
processor in the overlap zone.  When storing $n$-dimensional vectors such as
$x$ and $u$ above, each processor $p_i$ stores all the vector elements whose
indices fall in $J_i$.  Thus, vector coefficients in overlap zone columns will
be replicated in multiple processors. In the example
of~\cref{FIG:NonzeroPart}, a vector $x \in \reals^8$ will have the replication
pattern shown in~\cref{FIG:xCoefReplication}, with three overlap zones
corresponding to matrix columns $2$, $4$, and $6$.  The vector element $x_2$
is replicated twice, $x_4$ three times, and $x_6$ twice.

In this scheme, every processor receives the same number of matrix nonzeros,
give or take $1$, regardless of the sparsity pattern of $A$.  However,
$n$-vectors are represented in a possibly unbalanced manner.

\section{SpMVs with Nonzero Partitioning}
\label{SEC:SpMV}
Given our vector and matrix distribution scheme, we may calculate $Ax$ as
follows, using the terminology of MPI~\cite{MPIRef}:

\begin{enumerate}
\item \label{MG:STEP:Compq} Each processor $p_i$ locally calculates $A_{(i)} x_{(i)} = y_{(i)}$
\item \label{MG:STEP:Reduceq} We sum the vectors $y_{(i)}\in \realsm$ over all processors
using the \ttt{MPI\_Allreduce} operation. 
Every processor thus receives the vector $\sum_{i=0}^{P-1} y_{(i)} = y = Ax$.
\end{enumerate}

\begin{figure}
\begin{center}
\scalebox{0.4}{
\begin{tabular}[t]{c c c c c}
\begin{tabular}[t]{ c c| c}
 \cline{1-2} 
 \multicolumn{1}{|c}{}  		& \multicolumn{1}{c|}{}			&$\times$\\
 \multicolumn{1}{|c}{\color{red}*} 	& \multicolumn{1}{c|}{\color{red}*}	&\\
 \multicolumn{1}{|c}{}	 		& \cellcolor{gray!25}			&\\
 \multicolumn{1}{|c}{}  		& \cellcolor{gray!25}			&\\
 \multicolumn{1}{|c}{}  		& \cellcolor{gray!25}			&\\
 \multicolumn{1}{|c}{\color{red}*}	& \cellcolor{gray!25}			&\\
 \cline{1-2}
\end{tabular}

\begin{tabular}[t]{ |c|}
 \hline 
 \multirow{2}{*}{\color{red}$x_1$} 	\\
					\\
  \color{red}$x_2$			\\
 \hline
\end{tabular}

\begin{tabular}[t]{ c|c|}
 \cline{2-2}
  \multirow{2}{*}{$=$} & \multirow{2}{*}{\color{red}$y_{(0)}$} \\
  & 			\\
  &\cellcolor{gray!25} 	\\ 
  &\cellcolor{gray!25} \\
  &\tikzmark{Sp0} \cellcolor{gray!25} \\
  &\cellcolor{gray!25}  \\
 \cline{2-2}
\end{tabular} 

&

& 

\begin{tabular}[t]{|c|c}
 \hhline{-~} 
 \cellcolor{gray!25}	& $\times$\\
 \cellcolor{gray!25}	&\\
 			&\\
 *			&\\
 *			&\\
 *			&\\
 \cline{1-1}
\end{tabular}

\begin{tabular}[t]{|c|}
 \hline 
  \multirow{2}{*}{$x_2$} 	\\
				\\ 
 \hline
\end{tabular}

\begin{tabular}[t]{ c|c|}
 \hhline{~-}
  &\cellcolor{gray!25}  \\
  &\cellcolor{gray!25}  \\
  &		    	\\
  $=$ &$y_{(1)}$	\\
  & \tikzmark{Sp1}  \\
  &		    \\
 \cline{2-2}
\end{tabular} 

&

\begin{tabular}[t]{c}
 \\ 
 \\
 \\
 \\
 \\
 \\
  $\ldots$\\
\end{tabular} 

&

\begin{tabular}[t]{ c c c }
 \cline{1-2}
 \multicolumn{1}{|c}{} 		& \multicolumn{1}{c|}{} 		& $\times$ \\
 \multicolumn{1}{|c}{} 		& \multicolumn{1}{c|}{\color{green}*} 	&	\\
 \multicolumn{1}{|c}{} 		& \multicolumn{1}{c|}{}  		&	\\ 
 \multicolumn{1}{|c}{} 		& \multicolumn{1}{c|}{}  		& 	\\
 \multicolumn{1}{|c}{} 		& \multicolumn{1}{c|}{\color{green}*} 	&       \\
 \multicolumn{1}{|c}{\color{green}*} & \multicolumn{1}{c|}{} 		&	\\
 \cline{1-2}
\end{tabular}

\begin{tabular}[t]{ |c|}
 \hline 
 \multirow{2}{*}{\color{green}$x_7$}	\\
					\\
  \color{green}$x_8$			\\
 \hline
\end{tabular}

\begin{tabular}[t]{ c|c|}
 \cline{2-2}
  & \\
  & \\
  & \\ 
  $=$& \color{green}$y_{(6)}$ \\
  &\tikzmark{Sp6}  \\
  &		   \\
 \cline{2-2}
\end{tabular} 

\\
& & & & \\
& & & & \\
\multicolumn{5}{c}{ 
  \begin{tikzpicture}[remember picture]
  \node (Sabove) at (0.5,0.5){};
  \node (Sleft) at (0,0){};
  \node (Sright) at (1,0){};
  \node (Sbelow) at (0.5,-0.5){};
  \draw[line width=1em] (0,0) -- (1,0)(0.5,0.5) -- (0.5,-0.5);
  \end{tikzpicture}
}
\\
& & & & \\
& & & & \\

\begin{tabular}[t]{|c|}
 \hline
  \multirow{6}{*}{\color{red}\large{$y$}} \\
  \\
  \\ 
  \\
  \\
  \\
 \hline
\end{tabular}
\tikzmark{Sy0} 

& 

&

\begin{tabular}[t]{|c|}
 \hline
  \multirow{6}{*}{\large{$y$}} \\
  \\
  \\ 
  \\
  \\
  \\
 \hline
\end{tabular}
\tikzmark{Sy1}

&

\begin{tabular}[t]{c}
 \\
 \\
 \\ 
 \\
 \\
 \\
  $\ldots$\\
\end{tabular} 

&

\tikzmark{Sy6} 
\begin{tabular}[t]{|c|}
 \hline
  \multirow{6}{*}{\color{green}\large{$y$}} \\
  \\
  \\ 
  \\
  \\
  \\
 \hline
\end{tabular}
\end{tabular}
\begin{tikzpicture}[overlay, remember picture, shorten >=.5pt, shorten <=.5pt]
    \draw [->] ({pic cs:Sp0}) [bend right] to (Sleft);
    \draw [->] ({pic cs:Sp1}) [bend right] to (Sabove);
    \draw [->] ({pic cs:Sp6}) [bend left] to (Sright);
    \draw [->] (Sleft) [bend left] to ({pic cs:Sy0});
    \draw [->] (Sbelow) [bend left] to ({pic cs:Sy1});
    \draw [->] (Sright) [bend right] to ({pic cs:Sy6});
\end{tikzpicture}}
\end{center}
\caption{Nonzero partition SpMV, $Ax = y$.}\label{FIG:SparseReduce}
\end{figure}
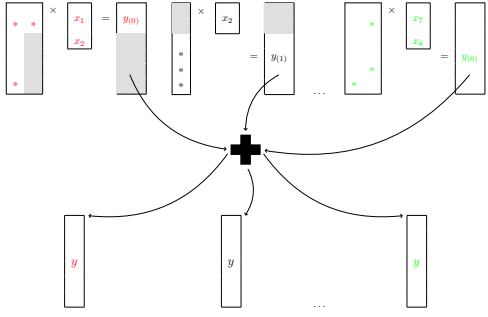

\cref{FIG:SparseReduce} illustrates this procedure. SpMVs with nonzero
partitioning are straightforward to implement and work in essentially the same
manner as they would with column partitioning. The complexity of the first
step is $\bigo(Z/P)$ and the complexity of the second step is $\bigo(m\log P)$
using standard parallel reduction algorithms.  Thus, the overall complexity is
$\bigo(Z/P + m \log P)$.

\section{SpVTMs with Nonzero Partitioning}
\label{SEC:SpVTM}
\subsection{Calculations}
We now consider calculations of the form $v\transpose A = u\transpose$.  If we
were to use a column-partitioned representation of $A$, then this operation
would require no communication, assuming that we are replicating the length-$m$
vector $v$ on all processors as we have already supposed.  In our
nonzero-based partitioning scheme, however, some communication is needed,
because overlap-zone columns are split between different processors.
Specifically, suppose that $j$ is 
the $k^{\text{th}}$ zone (counting from left to right)
and the storage for $A_j$ is
split among the processors in $S_k = \{p_{\ell},p_{\ell+1},\ldots,p_r\}$. 
Each processor $p \in S_k$ stores a 
partial column $A_j^p$ such that $A_j = \sum_{p\in S_k} A_j^p$. 
Globally, we need to compute $u_j = v\transpose A_j$, and then 
replicate this value among all processors serving the overlap zone. 
Locally, however, each processor is only 
able to compute $u_j^p = v\transpose A_j^p$, 
but if we sum these values we obtain
\[
\textstyle{
\sum_{p \in S_j}
v\transpose A_j^p = v\transpose\left(\sum_{p \in S_j} A_j^p\right) =
v\transpose A_j = u_j}.
\]
So, among the processors for each overlap zone, we simply need to sum the
scalars $u_j^p = v\transpose A_j^p$  and then broadcast the result.  We thus
arrive at the following algorithm:


\begin{enumerate}
 \item Each processor $p_i$ locally computes $v\transpose A_{(i)} = u_{(i)}$.  
 \item The processors $p \in S_k$ corresponding to each overlap zone $j_k$
 compute the sum $\sum_{p\in S_k} u_{j_k}^p$ and broadcast it throughout
 $S_k$.
\end{enumerate}

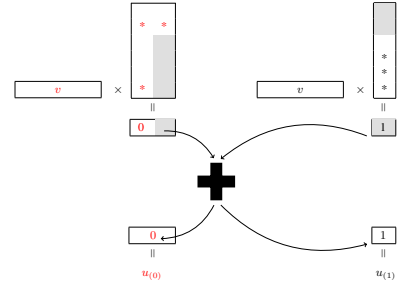
\begin{figure}
\begin{center}
\scalebox{0.5}{
\begin{small}
\begin{tabular}{c c c}

\begin{tabular}[b]{| c c c c c |}
 \hline 
 & &  \color{red} $v$ & & \\
 \hline
\end{tabular}

\begin{tabular}[b]{ c c | c|}
 \cline{2-3} 
	 & \multicolumn{1}{|c}{}  		&		\\
	 & \multicolumn{1}{|c}{\color{red}*} 	& \color{red}*	\\
	 & \multicolumn{1}{|c}{}  		& \cellcolor{gray!25}\\
	 & \multicolumn{1}{|c}{}  		& \cellcolor{gray!25}\\
	 & \multicolumn{1}{|c}{}		& \cellcolor{gray!25}\\
 $\times$& \multicolumn{1}{|c}{\color{red}*}	& \cellcolor{gray!25}\\
 \cline{2-3}
\end{tabular}

&  &

\begin{tabular}[b]{| c c c c c |}
 \hline 
 & & $v$ & &  \\
 \hline
\end{tabular}

\begin{tabular}[b]{ c|c|}
 \hhline{~-} 
	 &\cellcolor{gray!25}	\\
	 &\cellcolor{gray!25}	\\
	 &		\\
	 & *		\\
	 & *		\\
 $\times$& *	\\
 \cline{2-2}
\end{tabular}

\\

\multicolumn{1}{r}{\verteq $\;\;\;\;\;\:$ } & & \multicolumn{1}{r}{\verteq $\;\;\:$}

\\

\multicolumn{1}{r}{
  \begin{tabular}[b]{|m{0.25cm} m{0.1cm}|}
    \hline 
      \color{red} 0 & \cellcolor{gray!25} \tikzmark{STp0}  \\
    \hline
  \end{tabular} 
}

&  & 

\multicolumn{1}{r}{ \tikzmark{STp6}
  \begin{tabular}[b]{|m{0.2cm}|}
    \hline 
      \cellcolor{gray!25} 1 \\
    \hline
  \end{tabular}
}

\\
& & \\

& 
  \begin{tikzpicture}[remember picture]
  \node (STabove) at (0.5,0.5){};
  \node (STleft) at (0,0){};
  \node (STright) at (1,0){};
  \node (STbelow) at (0.5,-0.5){};
  \draw[line width=1em] (0,0) -- (1,0)(0.5,0.5) -- (0.5,-0.5);
  \end{tikzpicture}
&

\\

& & \\

 \multicolumn{1}{r}{
  \begin{tabular}[b]{c c}
    \hline 
      \multicolumn{2}{|m{0.8cm}|}{$\;\:\:$ \color{red} $0$ \tikzmark{STu0} }  \\
    \hline
  \end{tabular} 
}

&

& 

\multicolumn{1}{r}{ \tikzmark{STu6}
  \begin{tabular}[b]{|m{0.2cm}|}
    \hline 
      $1$ \\
    \hline
  \end{tabular}
}

\\

\multicolumn{1}{r}{\verteq $\;\;\;\;\;\:$ } & & \multicolumn{1}{r}{\verteq $\;\;\:$} 

\\

\multicolumn{1}{r}{\color{red} $u_{(0)}$ $\;\;\:$ } & & \multicolumn{1}{r}{$u_{(1)}$}

\\

\end{tabular}
\end{small}
\begin{tikzpicture}[overlay, remember picture, shorten >=.5pt, shorten <=.5pt]
     \draw [->] ({pic cs:STp0}) [bend left] to (STabove);
     \draw [->] ({pic cs:STp6}) [bend right] to (STabove);
     \draw [->] (STbelow) [bend left] to ({pic cs:STu0});
     \draw [->] (STbelow) [bend right] to ({pic cs:STu6});
\end{tikzpicture}}
\end{center}
\caption{Nonzero partition SpVTM calculation for overlap zone $j_0 = 1$ of the example, for which $S_0 = \{p_0,p_1\}$.}\label{FIG:SparseTransposeReduce}
\end{figure}

\cref{FIG:SparseTransposeReduce} illustrates this procedure for the first
overlap zone in the example above.  In general,
the local
matrix multiplication step has complexity $\bigo(Z/P)$, just as in the
previous algorithm.  In MPI, we implement the second step by using
\ttt{MPI\_Allreduce} operations on specialized communicators, one for each
overlap zone. 

\begin{table}[hb]
\caption{Overlap zones and associated processor sets for the example matrix of
         \cref{FIG:SparseMtx,FIG:NonzeroPart,FIG:xCoefReplication}.}
         \label{TB:Zones}
\begin{center}
\vspace{-1.5ex}
\begin{tabular}{|c|c|l|}
\hline
Zone       & Matrix       & \multicolumn{1}{c|}{Processor} \\
Rank $k$ & Column $j_k$ & \multicolumn{1}{c|}{Set $S_k$} \\
\hline
$0$ & $2$ & $S_0 = \{p_0,p_1\}$ \\
\hline
$1$ & $4$ & $S_1 = \{p_2,p_3,p_4\}$ \\
\hline
$2$ & $6$ & $S_2 = \{p_4,p_5\}$ \\
\hline
\end{tabular}
\end{center}
\end{table}

In general, we let $z$ be the number of overlap zones and denote them
$j_0 < j_2 < \cdots < j_{z-1}$, 
with respective associated processor sets $S_0,\ldots,S_{z-1}$.
\cref{TB:Zones} shows the zones and sets for the example matrix of
\cref{FIG:SparseMtx}. Only consecutive
$S_k$ sets can intersect, and only by one element. Such an intersection can
only occur when a processor owns more than one column, shares part of its
first column with the previous processor, and shares part of its last column
with the next processor.  In our example, this
situation occurs for zones $1$ and $2$ (respectively corresponding to matrix
columns $4$ and $6$), which both contain processor $p_4$.

As a consequence of this overlap pattern, the processor sets $S_0, S_2,
\ldots$ of the even overlap zones $j_0, j_2, \ldots$ must be disjoint, as are
the processor sets $S_1, S_3, \ldots$ of the odd overlap zones $j_1, j_3,
\ldots$.
Therefore, 
it should be possible,
assuming a sufficiently capable communication network, to compute 
all the reductions required for the even zones
simultaneously in time $\bigo(\log V) \subseteq \bigo(\log P)$, where $V =
\max_{k=1,\ldots,z} \{\lvert S_{j_k}\rvert\}$ is the maximum number of
processors associated with an overlap zone.  
The same observation applies to the odd zones, so the reductions 
may be performed in two
$\bigo(\log V)$ steps and overall complexity of the calculation is $\bigo(Z/P
+ \log V) \subseteq \bigo(Z/P + \log P)$.

In versions of MPI implementing the MPI-3 standard~\cite{MG:MPI3point1}, 
one can use
overlapping nonblocking reduction operations instead of the ``even-odd''
alternation described above, but the complexity of the reduction operations
remains $\bigo(\log V) \subseteq \bigo(\log P)$.

\subsection{Communicator Setup}
\label{SEC:Communicator}
Our SpVTM algorithm requires some one-time setup: specifically, we
need to create an MPI communicator for each overlap zone.  We now demonstrate
that the time required to set up the overlap zone communicators is relatively
insignificant: using appropriate parallel computations, they can be created in
$\bigo(\log P)$ time, which is of lower order than a single multiplication.

To accomplish this, we must be careful in our use of MPI primitives.
MPI contains the powerful, general communicator-creation primitive
\ttt{MPI\_Comm\_split}, but it uses all-to-all communication
operations and therefore has complexity $\Omega(P)$; we explicitly avoid such
operations.



Throughout this section, we consider the processors to be ordered from
``left'' to ``right'', that is, processor $p_{i-1}$ is considered to be to the
left of processor $p_i$, and processor $p_{i+1}$ to be on the right. 
At the beginning
of the procedure, each processor $p_i$, $i=1,\ldots,P-1$, determines the
index of its first column $j^{\trm{f}}_i \in J_i$ and the index of its last
column $j^{\trm{l}}_i \in J_i$; this step requires
only constant time if the matrix elements are stored in column major order.
%
%
The communicator setup procedure operates as follows:

\subsubsection{Determine overlap zones}

Using the primitive \texttt{MPI\_Sendrecv}, each processor $p_i$ (except the
last) sends $j^{\trm{l}}_i \in J_i$ to processor $p_{i+1}$. 
Conversely, also using
\ttt{MPI\_Sendrecv}, each processor $p_i$ (except the first) sends
$j^{\trm{f}}_i \in J_i$ to $p_{i-}1$. 
Then each processor $p_i$ computes
the variables \texttt{needLeft} and \texttt{needRight}, flags respectively
indicating whether a processor shares responsibility for its first column with
the previous processor, or responsibility for its last column with the next
processor. Specifically, 
\begin{align*} 
\ttt{needLeft}
&=
\left\{
\begin{array}{ll}
1, & \textrm{if~} j^{\trm{f}}_i = j^{\trm{l}}_{i-1}, \\
0, & \textrm{otherwise}
\end{array}
\right.
 i = 1, \ldots, P-1 
\\
\ttt{needRight}
&=
\left\{
\begin{array}{ll}
1, & \textrm{if~} j^{\trm{l}}_i = j^{\trm{f}}_{i+1}, \\
0, & \textrm{otherwise}
\end{array}
\right.
 i = 0, \ldots, P-2.
\end{align*}
We set \texttt{needLeft} to $0$ in $p_0$ and \texttt{needRight} to
$0$ in $p_{P-1}$.

\subsubsection{Determine ranks of overlap groups}
A processor's ``left group'' is the set of processors sharing responsibility
for its first column, and its ``right group'' is the set of processors sharing
responsibility for its last column. 
A processor may be in two groups if it
shares its first column and last column with other processors, but these
columns are not the same, as occurs for processor $4$ in the example above. As
already established, a processor cannot be in more than two groups.

We next assign a unique rank from $0$ to $z-1$ to each group ($z$ being the
number of overlap zones), progressing from the lowest to highest-ranked
processors.  In our example matrix of \cref{FIG:SparseMtx}, the ranks are as
shown in \cref{TB:Zones}.  Group $0$, with $S_0 = \{p_0,p_1\}$ is the right
group of $p_0$ and the left group of $p_1$.  Group $1$,
with $S_1 = \{p_2,p_3,p_4\}$ is the right group of $p_2$, both the
left and right group of $p_3$, and the left group of 
$p_4$.  Finally, group $2$, with $S_2 = \{p_4,p_5\}$, is the right group of
$p_4$ and the left group of $p_5$.

To create this ranking and identify all left and right groups,
each processor first computes an integer variable
\ttt{leftGroupEnd} that is $1$ if the processor has a left group and
is the highest-rank processor in that group; otherwise, it is $0$.  The value
assigned to \ttt{leftGroupEnd} on processor $p_i$ is
\[
(\ttt{needLeft} = 1) \wedge 
\big(
(\ttt{needRight} = 0) \vee (j^{\trm{f}}_i \neq j^{\trm{l}}_i)
\big).
\]




Next, we perform an additive forward scan (or parallel prefix operation,
see for example \cite[Chapter 4 and Appendix A]{MG:Blelloch90}) on
\texttt{leftGroupEnd}.  The resulting value is the number of complete groups
up to and including each processor's left group.  We call this value
\texttt{rightGroup}, because it is the rank, from $0$ to $z-1$, of the
processor's right group, if it has one (otherwise its value is immaterial).
We then calculate the rank of each processor's left group (if it has one), by
\begin{center}
\texttt{leftGroup = rightGroup - leftGroupEnd}.  
\end{center}

\subsubsection{Determine group extents}
Consider a segmented addition operator $\circ$ operating on pairs of
integers in the following manner:
\[
\left(
\begin{array}{cc} s \\ k \end{array}
\right)
\circ
\left(
\begin{array}{cc} t \\ l \end{array}
\right)
=
\left\{
\begin{array}{ll}
\left( \begin{array}{cc} s+t \\ l \end{array}\right),\quad & \text{if~} k=l \\ \, \\
\left( \begin{array}{cc} t \\ l \end{array}\right), & \text{if~} k \neq l.
\end{array}
\right.
\]
This operator is noncommutative, but easily shown to be associative; 
it is therefore suitable for parallel prefix operations.

We next perform a forward scan using the $\circ$ operator on the pairs
\[
\left(
\begin{array}{cc} \text{\texttt{needLeft}} \\ \text{\texttt{leftGroup}} \end{array}
\right).
\]
In each processor $p_i$, let \texttt{procsOnLeft} denote the first element of
the result; it is the number of lower-ranked processors in the current
processor's left group. Next, we perform a backward scan, using the same
operator $\circ$, on the pairs
\[
\left(
\begin{array}{cc} \text{\texttt{needRight}} \\ \text{\texttt{rightGroup}} \end{array}
\right).
\]
In each processor, let \texttt{procsOnRight} denote the first element of the
result; this value is the number of higher-ranked processors in the current
processor's right group.

\begin{table}[tbp]
\caption{Communicator setup variable values for 
         matrix in~\cref{FIG:SparseMtx}.}\label{FIG:Variables}
\begin{center}
\begin{tabular}{|c|c|c|c|c|c|c|c|}
\hline
 & $p_0$ &$p_1$ &$p_2$ &$p_3$ &$p_4$ &$p_5$ &$p_6$ \\
\hline
\ttt{needLeft} & 0 & 1 & 0 & 1 & 1 & 1 & 0\\
\hline
\ttt{needRight} & 1 & 0 & 1 & 1 & 1 & 0 & 0\\
\hline
\ttt{leftGroupEnd} & 0 & 1 & 0 & 0 & 1 & 1 & 0\\
\hline
\ttt{rightGroup} & 0 & 1 & 1 & 1 & 2 & 3 & 3 \\
\hline
\ttt{leftGroup} & 0 & 0 & 1 & 1 & 1 & 2 & 3\\
\hline
\ttt{procsOnLeft} & 0 & 1 & 0 & 1 & 2 & 1 & 0 \\
\hline
\ttt{procsOnRight} & 1 & 2 & 2 & 1 & 1 & 0 & 0 \\
\hline
\end{tabular}
\label{TB:Variables}
\end{center}
\end{table}

Each processor now has sufficient information to determine the full extent of
its left and right groups (which may be the same group, or empty).
\cref{TB:Variables} shows all the calculations for the example matrix of
\cref{FIG:SparseMtx}; note that value of \texttt{procsOnLeft} is unused in
processors for which $\texttt{needLeft} = 0$ and similarly the value of
\texttt{procsOnRight} is unused in processors for which $\texttt{needRight} =
0$.

\subsubsection{Create communicators}
Each processor now uses \ttt{MPI\_Group\_range\_incl} to create MPI
group objects corresponding to its left and right groups; if these groups are
the same, only one MPI group object is created. Every processor now
simultaneously
calls \texttt{MPI\_Comm\_create} with the group argument set to the MPI
group object for the even-ranked group it participates in, if any.  Each
processor can be in at most one even-ranked group.  If it is not in such a
group, the processor supplies the argument \texttt{MPI\_GROUP\_NULL}. This
operation simultaneously creates all even-numbered groups.
Next, in a similar manner, we simultaneously create all odd-numbered groups.
This operation completes the communicator setup.

To summarize the complexity of the operations above, all the numeric
calculations require constant time, while the scans require $\bigo(\log P)$
time.  The MPI groups created may each be described by a single
range-stride triplet, so their creation needs only constant time. Finally,
creating an MPI communicator requires $\bigo(\log S)$ time, where $S$ is
the size of the communicator.  Therefore, simultaneous creation of the even
communicators requires $\bigo(\log V) \subseteq \bigo(\log P)$ time, and the
same goes for 
the
creation of the odd communicators. Therefore, the
entire communicator creation process has complexity $\bigo(\log P)$.

\section{Implementation Details}
\label{SEC:Implementation}
We created a \ttt{C++} implementation consisting of two driver programs
\ttt{ColP} and \ttt{NzP}, a class implementing our nonzero partitioning
techniques, and a standard local sparse-matrix-multiply kernel
\ttt{default\_dcscmv}. The \ttt{ColP} driver implements the column
partitioning technique: each processor reads in a contiguous span of the
columns of the matrix stored in Intel \ttt{MKL}'s \ttt{BLAS CSC}
format~\cite{MG:IntelCsc}, and then performs $1{,}000$ wraps of SpMV-SpVTM
pairs. This pattern is meant to reflect that algorithms involving wide or tall
matrices must typically perform equal numbers of matrix-vector and
vector-transpose-matrix multiplications. We implement the local matrix-vector
multiplies with the
\ttt{default\_dcscmv} kernel, and the sums needed by the SpMV operations with
\ttt{MPI\_Allreduce} reduction primitives.

The \ttt{NzP} driver implements the nonzero partitioning approach. For this technique, we store the matrix as a stream of \ttt{(row, column,
value)} triples in column major order, and each processor core reads in a
contiguous span of that data. After reading in the data, we use our nonzero
partitioning class to determine the overlap zones and creates the necessary
communicators as described in~\cref{SEC:Communicator}. Then, similarly to
\ttt{ColP}, \ttt{NzP} performs $1{,}000$ wraps of SpMV-SpVTM pairs.
The multiplies consists of local sparse matrix-vector and
vector-transpose-matrix multiplies, again executed by the kernel
\ttt{default\_dcscmv}, and MPI reduction operations, which include
reductions performed across overlap zones in the case of the SpVTM
calculations.

\section{Experimental Results}
\label{SEC:Comp}
Using an HPC system~\cite{MG:Caliburn} with Xeon e5-2694v4 processors, 
the Intel Omni-Path
interconnect fabric, the GNU C++ compilers, and MPICH, we tested both approaches with a highly unstructured wide
matrix obtained from 
the UCI ``Twenty Newsgroups''
(News20) data set~\cite{MG:Keerthi05:News20, ML}. The problem's 
document-term matrix $A$ has
19,996 rows, 1,355,191 columns and a total of 9,097,916 nonzero coefficients.
While the average density is approximately $7$ nonzeros per column, the matrix
has $25$ columns having at least 9,998 nonzero elements, with the $10$ densest
columns ranging between 11,872 and 18,531 nonzeros. 

\begin{figure}[htb]
\begin{center}
\includegraphics[scale=0.4]{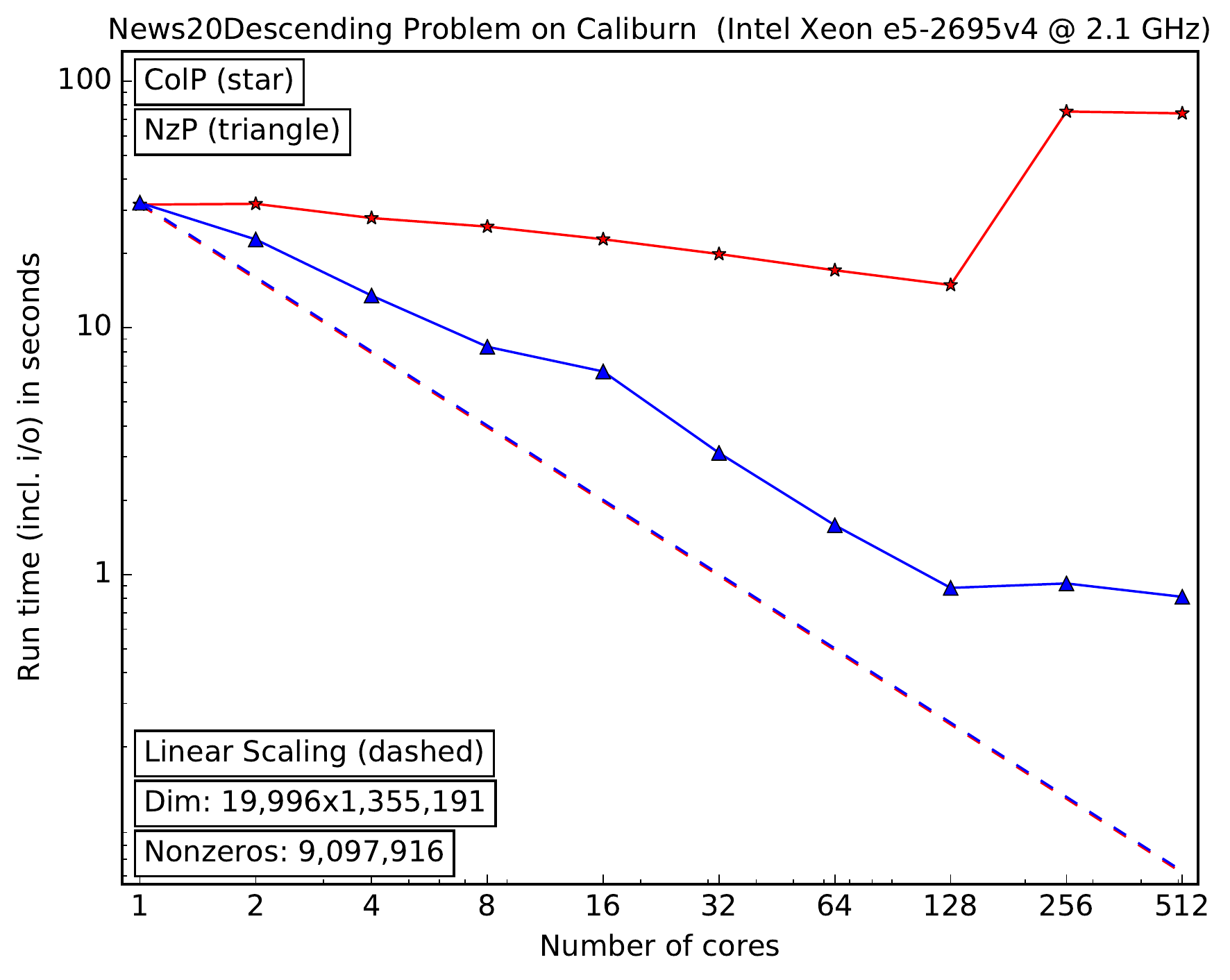} 
\end{center}
\vspace{-4ex}
\caption{Strong scaling graph for News20, descending density column order
(log-log).}\label{FIG:DescScaling}
\end{figure}

For our first set of tests, 
we arranged the
columns of $A$ in descending order based on their density, which is the
worst-case scenario for the column partitioning approach.
\Cref{FIG:DescScaling} is a strong scaling graph for both techniques, 
with the number of processor cores appearing on the horizontal
axis and the program running time, including I/O time, on the vertical axis. 
Clearly, the \ttt{NzP} algorithm exhibits much better scaling than \ttt{ColP},
since it does not have to contend with workload imbalances between processor
cores.  \ttt{NzP} exhibits good scaling behavior through $128$ processor
cores, after which little further speedup is obtained, however, at that point,
the total computation time is less than $1$ second for $1{,}000$
multiplication pairs. \Cref{TB:ImbalanceDesc} shows the nonzero imbalance
(expressed as a percentage) for \ttt{ColP} and the number of overlap zones for
\ttt{NzP}. The nonzero imbalance ($\Delta$) is calculated by dividing the
difference between the maximum ($\Xi$) and minimum ($\xi$) number of nonzeros
assigned to a single processor core by the average number of nonzeros per
core ($Z / P$), \emph{i.e.} $ \Delta = (\Xi - \xi) / (Z / P) = P(\Xi -
\xi) / Z$.

\begin{figure}[htb]
\begin{center}
\includegraphics[scale=0.4]{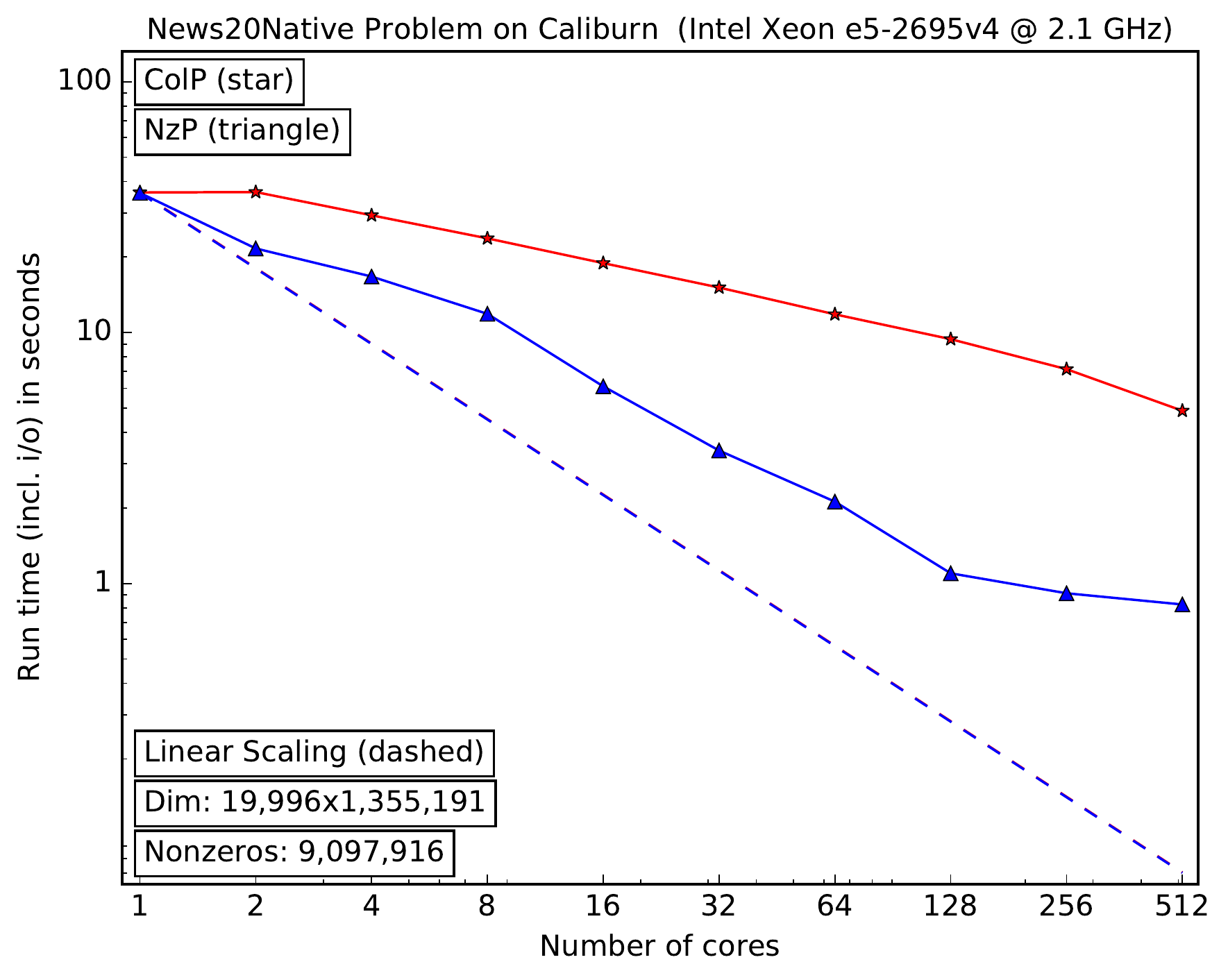}
\end{center}
\vspace{-4ex}
\caption{Strong scaling graph for News20, native column order
(log-log).}\label{FIG:NativeScaling}
\end{figure}

\Cref{FIG:NativeScaling} shows the performance of both methods on the News20
dataset with its columns in the ``native'' order as obtained from the UCI
repository.   Although this order is no longer the worst possible for column
partitioning, \Cref{TB:ImbalanceNative} indicates that the resulting imbalance
is very substantial.  As a result, the \ttt{NzP} code still performs
significantly better than the \ttt{ColP}.

Finally, we tested both approaches on a $2{,}000 \times 100{,}000$ randomly
generated matrix generated by the following procedure:
\begin{enumerate}
\item Based on initial density ($\rho$) and imbalance ($\iota^-, \iota^+$)
parameters, we compute respective lower and upper bounds
\begin{align*}
l &= \flo{\rho m} - \iota^- &
u &= \cel{\rho m} + \iota^+
\end{align*}
on the number of nonzeros per column.
\item Randomly generate the number of nonzeros for each column from the
uniform distribution on the interval $[l,u]$.
\item Specify row indices using a shuffle procedure assuring that all rows
have at least one nonzero element.
\end{enumerate}
 
\begin{figure}[htb]
\begin{center}
\includegraphics[scale=0.4]{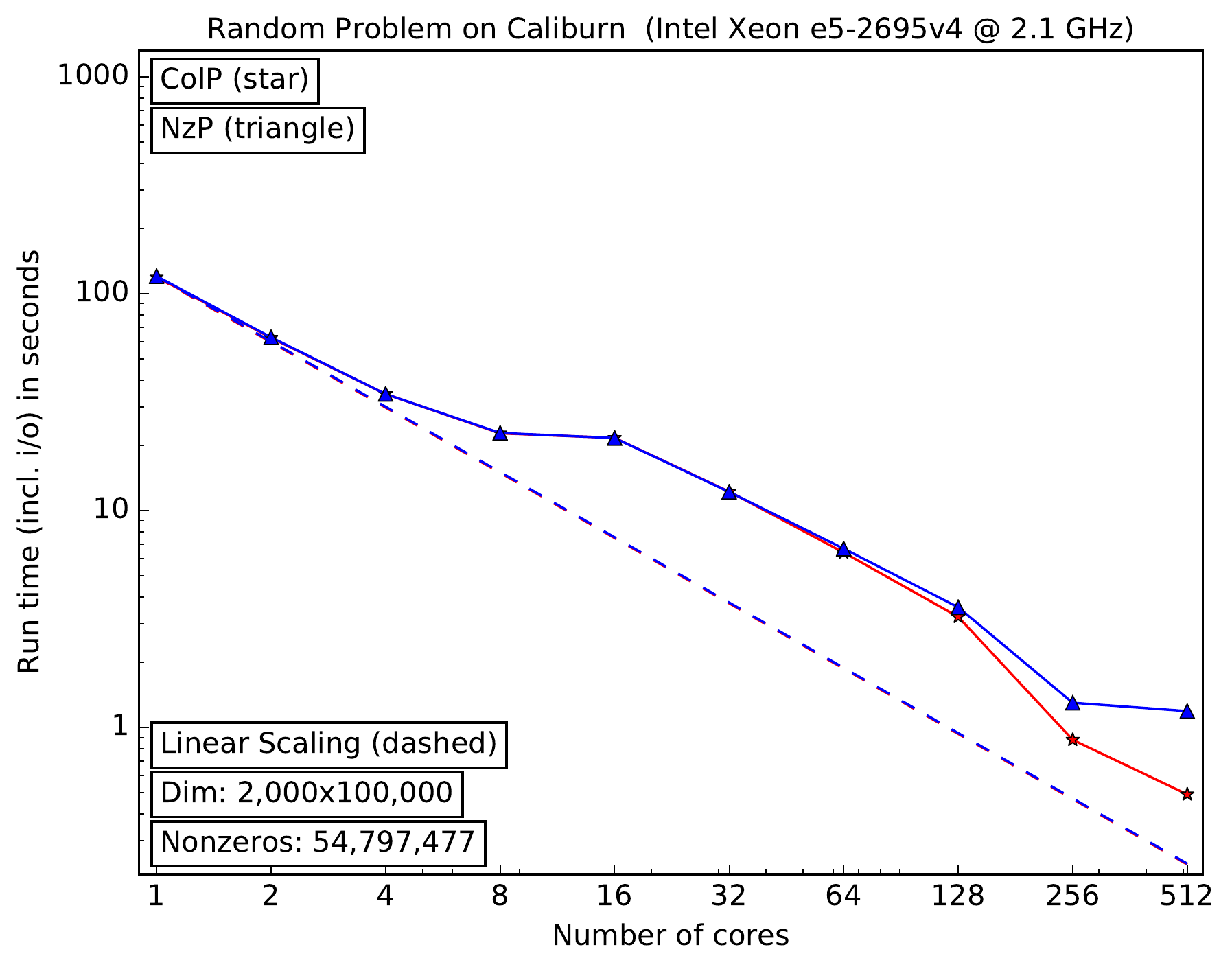}
\end{center}
\vspace{-4ex}
\caption{Strong scaling graph for a random sparse matrix
(log-log).}\label{FIG:RandomScaling}
\end{figure}

This matrix (denser than News20) has
$54{,}797{,}477$ nonzeros nearly evenly balanced between columns and requires a
large number of overlap zones for
\ttt{NzP}, as illustrated in \cref{TB:ImbalanceRandom}.  This situation is
essentially the best case for column partitioning.  As can be seen
from~\cref{FIG:RandomScaling}, the codes achieve essentially
identical scaling through $128$ cores, after which \ttt{NzP}'s scaling starts
to degrade faster than \ttt{ColP}'s. However, at that point, the total
computation time is approximately $1$ second.

\begin{table}[htbp]
\caption{Imbalance and overlap zones for the News20 matrix, descending density
order.}
\begin{center}
\begin{tabular}{|c|c|c|c|c|c|c|c|}
\hline
		& \ttt{ColP} & \ttt{NzP} \\
Processor cores & Imbalance \%  & Overlap Zones \\
\hline
$1$  		& 0\% & 0 \\
\hline
$2$ 		& 144\% & 1 \\
\hline
$4$ 		& 272\% & 3 \\
\hline
$8$ 		& 498\% & 6 \\
\hline
$16$	 	& 898\% & 12 \\
\hline
$32$ 		& 1,585\% & 27 \\
\hline
$64$ 		& 2,758\% & 54 \\
\hline
$128$ 		& 4,735\% & 111 \\
\hline
$256$ 		& 7,995\% & 186 \\
\hline
$512$ 		& 13,287\% & 408 \\
\hline
\end{tabular}
\label{TB:ImbalanceDesc}
\end{center}
\end{table}


\begin{table}[htbp]
\caption{Imbalance and overlap zones for the News20 matrix, native order.}
\begin{center}
\begin{tabular}{|c|c|c|c|c|c|c|c|}
\hline
		& \ttt{ColP} & \ttt{NzP} \\
Processor cores & Imbalance \% & Overlap Zones \\
\hline
$1$  		& 0\% & 0 \\
\hline
$2$ 		& 107\% & 1 \\
\hline
$4$ 		& 207\% & 3 \\
\hline
$8$ 		& 374\% & 6 \\
\hline
$16$	 	& 659\% & 13 \\
\hline
$32$ 		& 1,124\% & 27 \\
\hline
$64$ 		& 1,840\% & 55 \\
\hline
$128$ 		& 2,882\% & 111 \\
\hline
$256$ 		& 3,932\% & 218 \\
\hline
$512$ 		& 4,633\% & 435 \\
\hline
\end{tabular}
\label{TB:ImbalanceNative}
\end{center}
\end{table}


\begin{table}[htbp]
\caption{Imbalance and overlap zones for the Random matrix.}
\begin{center}
\begin{tabular}{|c|c|c|c|c|c|c|c|}
\hline
		& \ttt{ColP} & \ttt{NzP} \\
Processor cores & Imbalance \% & Overlap Zones \\
\hline
$1$  		& 0.00\% & 0 \\
\hline
$2$ 		& 0.02\% & 1 \\
\hline
$4$ 		& 0.48\% & 3 \\
\hline
$8$ 		& 0.98\% & 7 \\
\hline
$16$	 	& 2.35\% & 15 \\
\hline
$32$ 		& 3.33\% & 31 \\
\hline
$64$ 		& 6.62\% & 63 \\
\hline
$128$ 		& 8.79\% & 127 \\
\hline
$256$ 		& 13.24\% & 255 \\
\hline
$512$ 		& 22.50\% & 510 \\
\hline
\end{tabular}
\label{TB:ImbalanceRandom}
\end{center}
\end{table}

In summary, we conclude that \ttt{NzP} is relatively insensitive the sparsity
pattern of the matrix, as expected.  It is far more robust than
column partitioning, performing much better for unbalanced sparsity patterns
but matching column partitioning's performance on well balanced matrices,
except at the very highest processor counts.

\section{Final Remarks}
\label{SEC:Final}
First-order numerical algorithms involving sparse matrices typically alternate
between matrix-vector multiplications and simple vector operations such as
addition, scaling, and inner products.  As illustrated in
\cref{FIG:xCoefReplication}, the nonzero partitioning approach causes the
workload of such simple vector operations to become somewhat unbalanced,
because the number of stored vector coefficients varies between processors.
The worst possible case for these simple operation occurs when one processor
owns $\cel{Z/P}$ columns, each with a single nonzero.  In this case, local
vector operations have complexity $\bigo(Z/P)$ and inner products have
complexity $\bigo(Z/P + \log P)$.  In balanced column partitioning, the
respective complexity of these operations is $\bigo(n/P)$ and $\bigo(n/P +
\log P)$.  With column partitioning, however, the complexity of SpMV and SpVTM
operations could be respectively as bad as $\bigo(nm/P + m\log P)$ and
$\bigo(nm/P)$, as compared to $\bigo(Z/P + m\log P)$ and $\bigo(Z/P + \log P)$
for nonzero partitioning.  

If we suppose that there is a constant bound on the ratio between the number
of matrix multiplication and 
the number of 
other operations, as is typically the case, then
the effect of using nonzero partitioning is to improve the complexity of the
dominant matrix multiplication operations at the cost of worsening the
complexity of some nondominant operations.  The highest complexity
non-multiplication operation is the calculation of inner products, which
becomes $\bigo(Z/P + \log P)$, the same as the SpVTM operation and lower than
the SpMV operation.  Therefore, while non-multiplication operations become
more time consuming, they cannot become dominant in worst-case complexity
terms.
Meanwhile, the multiplication operations, which typically dominate running
time, generally become more balanced than in the column partitioning approach,
and remain so even for pathological sparsity structures.  In general, we are
trading off far better load balance in dominant operations for somewhat
inferior load balancing of non-dominant operations. We have used the nonzero
partitioning technique to implement various first-order optimization methods,
such as spectral gradient and conjugate gradient, and have observed that the
extra time required for simple vector operations does not have a significant
impact on overall computation time.

A natural topic for further research is whether similar techniques may be used
when $m$ and $n$ are of comparable magnitudes and both $m$-vectors and
$n$-vectors are stored in a distributed manner.  Such situations appear to be
considerably more complicated.

\section*{Acknowledgments}
This research was supported in part by National Science 
        Foundation grant CCF-1115638.
This research used resources from the Rutgers Discovery Informatics Institute
{(RDI\textsuperscript{2})}, a user facility supported by the Rutgers
University Office of Research and Economic Development
(ORED)~\cite{MG:Caliburn}.
The authors acknowledge the Texas Advanced Computing Center (TACC) at The
University of Texas at Austin for providing HPC resources that have
contributed to the research results reported within this paper. See
\url{http://www.tacc.utexas.edu}.

\printbibliography

\end{document}